\documentclass[twocolumn,aps]{revtex4}
\usepackage{epsfig,graphicx}
\begin{document}

\title{\bf The stability of strained H:Si(105) and H:Ge(105) surfaces}

\author{Cristian V. Ciobanu\footnote{Author to whom correspondence should be addressed.
Email: cciobanu@mines.edu} and Ryan M. Briggs}

\affiliation{Division of Engineering, Colorado School of Mines,
Golden, CO 80401}

\begin{abstract}
We report atomic scale studies of the effect of applied
strain and hydrogen environment on the reconstructions of the
(105) Si and Ge surfaces. Surface energy calculations for
monohydride-terminated (001) and (105) reconstructions reveal that
the recently established single-height rebonded model is unstable
not only with respect to (001), but also
in comparison to other monohydride (105) structures. This finding
persists for both Si and Ge, for applied biaxial strains from -4\%
to 4\%, and for nearly the entire relevant domain of the chemical
potential of hydrogen, thus providing an explanation for the
recently observed H-induced destabilization of the Ge(105)
surface.
\end{abstract}

\maketitle

The epitaxial system Ge/Si(001) has been the focus of intense
investigations for more than two decades, acting both as a test
bed for our fundamental understanding of the strained-layer
growth, as well as a technological launching pad for promising
optoelectronic devices based on Ge/Si quantum-dots. While the
understanding of the formation of quantum dots (islands) has
progressed rapidly \cite{chemrev}, the desire to further diversify
and control growth morphologies has triggered studies of Ge/Si
epitaxy in the presence of other species. In particular, hydrogen
was shown to have a surfactant effect on the deposition of Ge on
low-index silicon surfaces: the island formation was suppressed in
the presence of atomic hydrogen, with the growth switching to the
layer-by-layer mode \cite{growth-si-ge-h,
growth-si-ge-h001suppress}. In the H-mediated Ge/Si(001) epitaxy,
Si atoms tend to segregate at the surface, and their exchange with
Ge atoms is reversible upon hydrogen desorption
\cite{H-Ge-Sisegre-001-lagally}. Recently, experiments have broken
ground in a different direction to address the influence of
hydrogen on high-index epitaxial systems. Fujikawa and coworkers
used scanning tunneling microscopy and electron energy loss
spectroscopy to investigate the H adsorption on Ge/Si(105), and
demonstrated the destabilizing effect of hydrogen on the surface
\cite{prl-H105-jap-2005}.

Motivated by these compelling experiments
\cite{prl-H105-jap-2005}, we have studied the influence of applied
biaxial strain and chemical potential of hydrogen on the surface
energy of (105) and (001) reconstructions of Si and Ge. Our model
potential calculations predict that for a wide range of applied
strain and H chemical potential, the single-height rebonded
\cite{kds, prl-105-jap-2002,prl-105-ita-2002,apl-105-shenoy-2002}
structure with monohydride termination has higher surface energy
than the H-terminated (001). Interestingly, this rebonded-step
structure (RS \cite{prl-105-jap-2002} or SR
\cite{apl-105-shenoy-2002}) also becomes unstable with respect to
the single-height unrebonded (SU) model that was originally
proposed for the configuration of (105) hut facets \cite{mo}.

\begin{figure}
  \begin{center}
   \includegraphics[width=7.0cm]{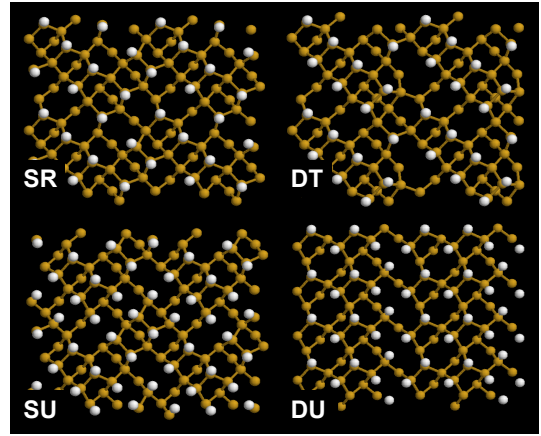}
  \end{center}
  \caption{(Color online) Monohydride (105) reconstructions with
  two different densities of hydrogenated bonds ({\em Hb}), 1.57{\em Hb}/$a^2$
  (SR, DT) and 2.35{\em Hb}/$a^2$ (SU, DU), where $a$ is the bulk lattice constant
  of Si or Ge.}
\label{H105models}
\end{figure}

In addition to SU and SR, we have studied all other (105)
structures \cite{ss-105-ciobanu-2003,prb-105-ciobanu-2004}
although, for brevity, we report results only for the SU, SR, DT,
and DU models (depicted in Fig.~\ref{H105models}). Using the
Tersoff potential \cite{si-ge} parameterized for the Si-H and Ge-H
systems \cite{si-h,ge-h}, we have performed full relaxations of
150 \AA-thick periodic slabs with either clean or H-passivated
reconstructions under applied biaxial strains $\epsilon$ in the
range $-4\% \leq \epsilon \leq 4\%$. With the dependencies on
strain ($\epsilon$) and H chemical potential ($\mu_H$) made
explicit, the surface energy $\gamma$ of a hydrogenated
reconstruction has been calculated as
\begin{equation}
\gamma(\epsilon, \mu_H) = (E(\epsilon)- N \mu (\epsilon) - N_H
\mu_H)/A(\epsilon),\label{equ}
\end{equation}
where $E$ is the total energy of the slab of area $A$, $N$ is the
number of Si or Ge atoms, $N_H$ is the number of H atoms
passivating the surface, and $\mu$ is the chemical potential (bulk
cohesion energy) of the Si or Ge atoms. Guided by experiments
\cite{prl-H105-jap-2005}, we have only considered here monohydride
terminations and estimated the range $\mu_H$ for which the surface
hydrogenation becomes thermodynamically favorable. We found that
the lowest value $\mu_H$ at which an H-passivated surface becomes
favorable over the pristine surface is only weakly dependent on
the reconstruction and strain, so we use one chemical potential
range for each material: $-3.1$ eV $\leq \mu_H\leq$ 0.0 eV for Si
and $-2.8$ eV $\leq \mu_H\leq$ 0.0 eV for Ge
\cite{Hsi001-northrup}.
\begin{figure}
  \begin{center}
   \includegraphics[width=7.5cm]{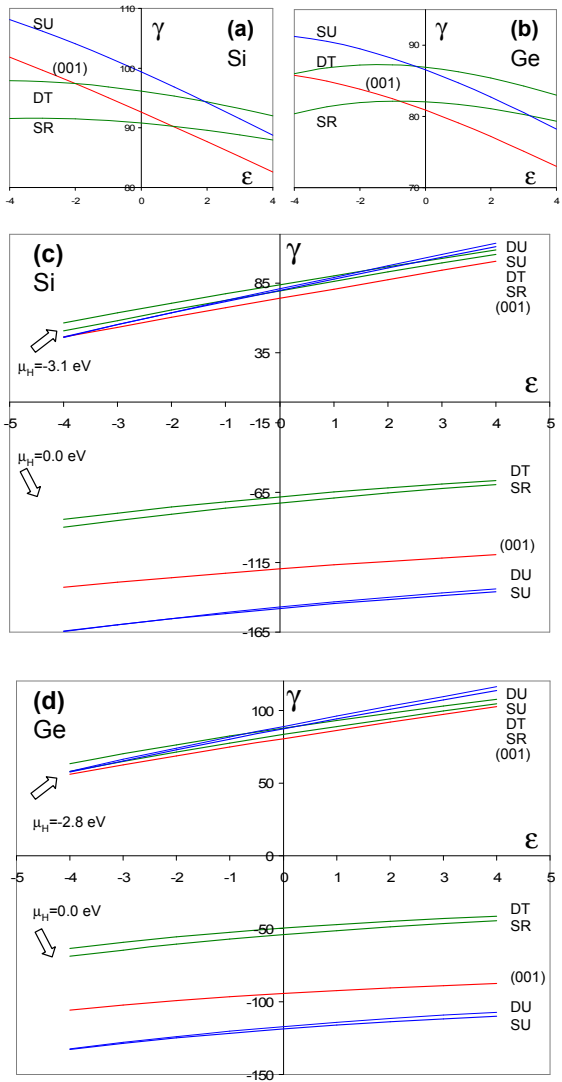}
  \end{center}
  \caption{(Color online) Surface energies (meV/\AA$^2$) of clean (a,b) and  monohydride (c,d)
  Si and Ge surfaces plotted versus biaxial strain $\epsilon$ (\%). The bottom panels
  show surface energy curves at the upper and lower bounds of the H chemical potential range
  for Si (c) and for Ge (d).
  } \label{clean-vs-H-abcd}
\end{figure}

Our main findings are summarized in Fig.~\ref{clean-vs-H-abcd}
which shows the energies of Si and Ge surfaces as functions of the
biaxial strain $\epsilon$. The results for clean surfaces
[Fig.~\ref{clean-vs-H-abcd}(a,b)] are consistent with previous
studies \cite{apl-105-shenoy-2002,
ss-105-ciobanu-2003,ss-Ge105-ita-2004} which point out that under
compressive strain the SR reconstruction becomes more stable than
the flat (001) surface.  In the case of hydrogenated Si and Ge
surfaces we find that:

(a) the ordering of surface energies corresponding to SR-(105) and
(001)-$2\times1$ is reversed upon hydrogenation for the entire
range of $\mu_H$ and $\epsilon$, for both Si and Ge.

(b) different (105) reconstructions have surface energies whose
plots vs. $\epsilon$ are grouped (bunched) according to the number
of hydrogenated bonds per area (e.g., SR and DT, SU and DU).

(c) the slope $\partial \gamma /\partial \epsilon$ is positive for
all monohydride reconstructions, irrespective of what its sign was
for clean surfaces.

On general grounds, observation (a) is consistent with ab initio
calculations which estimate that the average binding energy of H
on Ge/Si(105) is 0.1 eV smaller than that of H on Ge(001)
\cite{prl-H105-jap-2005}. Still, this estimate may not translate
readily into a surface energy difference (or ratio) between the
monohydride terminated SR and (001). We therefore pursue further
the origins of the reversal of energetic ordering of SR and (001)
upon hydrogenation. The ab-initio binding energy of the Ge-H bond
is 2.83 eV \cite{ge-h}, somewhat smaller than that of the Si-H
bond. Since this binding energy is an order of magnitude larger
than the difference \cite{prl-H105-jap-2005} in average binding of
H on Ge/Si(105) and H on Ge(001), the key factor in determining
the energetic ordering of monohydride reconstructions is the {\em
number of passivated bonds per area}. Indeed, if we focus on
surface energies at $\epsilon=0$\% and $\mu_H=0.0$ eV, we find
that their relative ratios can be largely accounted for by the
ratios of the areal density $\rho=N_H/A$ of monohydride bonds, as
shown in Table~\ref{table} for both Si and Ge. Thus, the SR
structure becomes unstable with respect to (001) mainly because it
has a lower density of H-bonds ({\em Hb}) at the surface (i.e.,
1.57{\em Hb}/$a^2$  as opposed to  2.00{\em Hb}/$a^2$ for (001),
where $a$ is the bulk lattice constant of either Si or Ge).
\begin{table}
\begin{center}%
\caption{Surface energy ratios of monohydride reconstructions at
$\epsilon=0$ and $\mu_H$=0.0 eV, compared to the ratios of
densities $\rho$ of passivated bonds. }
\begin{tabular}{l c c c c}
\hline \hline
              \   & SU/(001)   &  (001)/SR  & SU/SR & SR/DT \\
\hline %
$\gamma$ ratio, Si  & 1.24    & 1.65    & 2.04    & 1.07 \\
$\gamma$ ratio, Ge  & 1.26    & 1.75    & 2.20    & 1.09 \\
$\rho$ ratio        & 1.18    & 1.27    & 1.50    & 1.00 \\

\hline \hline
\end{tabular}
 \label{table}
\end{center}
\end{table}

Continuing the discussion at $\mu_H=0.0$ eV, when the H-bond
density $\rho$ is the same for two different reconstructions, then
their energetic ordering is dictated by subtler differences in
surface stress and in adsorption (binding) energy of H. In
general, these factors have less influence on surface energies
than the H-bond density (Table~\ref{table}), which is why, e.g.,
SR and DT curves are bunched together and are relatively far from
the other surface energy curves at $\mu_H=0.0$ eV
(Fig.~\ref{clean-vs-H-abcd}). We have found this bunching trend
[observation (b) above] for all the (105) reconstructions
\cite{prb-105-ciobanu-2004} that have the same density of
passivated bonds.

Variations of $\mu_H$ in the stated ranges lead, via
Eq.~(\ref{equ}), to variations of the energetic separation between
same-$\rho$ bunches, but leave the surface energy gaps within each
bunch unchanged (e.g., the gap between SR and DT does not depend
on $\mu_H$).  Because the lower bounds of $\mu_H$
(Fig.~\ref{clean-vs-H-abcd}) necessarily have magnitudes that are
similar to the Si-H or Ge-H binding energies, the separation
between equal-$\rho$ bunches can be wiped out in those limits
[refer, e.g., to the curves corresponding to $\mu_H=-3.1$ eV in
Fig.~\ref{clean-vs-H-abcd}(c)]. On the other hand, an increase of
$\mu_H$ by 0.3--0.5 eV above the lower bounds of $\mu_H$ is
sufficient for the bunches to be clearly distinguishable, thus for
the density of passivated bonds to set in as the dominant factor
for surface energetics.

We now turn to discussing the observation (c) listed above. The
positive slope of the surface energies in
Fig.~\ref{clean-vs-H-abcd}(c,d) is directly related to an increase
in the diagonal components of the surface stress of upon
hydrogenation. We have calculated the surface stress components
for all clean and monohydride Si (001) and (105) models and found
that each of the two diagonal components $\sigma_{11}$ and
$\sigma_{22}$ increases by as much as 40.0--65.0meV/\AA$^2$ upon
hydrogenation. In the case of Ge surfaces, the trend is the same
but the increases are smaller, in the range of 15.0--25.0
meV/\AA$^2$. Noting that stress components do not depend on
$\mu_H$, the increase of $\sigma_{11}$ and $\sigma_{22}$ is caused
by the stretching of surface bonds. Indeed, we have verified for
Si surfaces that the length of the dimer bonds is stretched by
about 2.5\% upon passivation, while the bridge-bonds made by the
rebonded atoms is increased by about 1.6\%. On Ge surfaces, dimers
are stretched by 2.0\% and bridges by 0.8\% , which explains the
smaller increase of stress components calculated for Ge.  Ab
initio calculations on H:Ge/Si(105) indicate that the bridge-bonds
actually decrease by 1.5\% \cite{prl-H105-jap-2005} for the
Ge/Si(105) system, which is also consistent with the
smaller stress variations 
for Ge surfaces compared to Si ones.

Before concluding, we comment as to why the empirical potentials
can capture the main energetic trends for hydrogenated surfaces.
We note that Tersoff \cite{si-ge} potentials have already been
found to predict the correct lowest-energy reconstruction and
strain behavior in the absence of hydrogen
\cite{prl-105-ita-2002,apl-105-shenoy-2002,ss-105-ciobanu-2003,ss-Ge105-ita-2004},
although most surface energy gaps were overestimated. For
passivated surfaces, the hybridization state of all atoms is
$sp^{3}$ \cite{prl-H105-jap-2005}, and thus major departures from
the fitting databases of these potentials do not occur. Subtle
electronic effects are therefore unlikely to dominate the energy
ordering of various monohydride reconstructions, which renders
them tractable at the level of Tersoff potentials \cite{si-ge,
si-h,ge-h}. We hope that the results obtained here will be
followed by further studies at the ab initio level, and we expect
quantitative improvements to emerge mainly for comparisons between
clean and H-passivated surfaces.

In conclusion, we investigated the relative stability of strained
(105) reconstructions with monohydride terminations. On purely
thermodynamic grounds, our calculations suggest that the
monohydride H:Ge/Si(105)-SR structure \cite{prl-H105-jap-2005} is
unstable for any Ge or Si-Ge coverage (or applied strain), not
only for coverages below a threshold value. However, the kinetic
barriers against SR decaying into other monohydride
reconstructions and the compression in the epitaxial layer make
the H:Ge/Si(105) experimentally observable as reported in
Ref.~\cite{prl-H105-jap-2005}. Since the SU model has the largest
possible surface density of monohydrides and a structure based on
single-height steps, it has been found here to also have the
lowest surface energy for the relevant ranges of strain and H
chemical potential. Thus, it is conceivable that the SU model may
be confirmed experimentally by pursuing different surface
preparation recipes that diminish the influence of kinetics while
avoiding the formation of dihydrides. The results presented here
could also be relevant for future experiments on the Si(105)
surface, which has long been known to be rough and disordered on
atomic scales \cite{prl-105-jap-2002,tomitori}. If maximal
monohydride coverage can be achieved on Si(105), the present
results indicate that Si(105) would become ordered in the presence
of H, with a reconstruction given by the SU model.

{\em Acknowledgments.} The authors gratefully acknowledge the
support of NCSA through grant no. DMR-050031N. RMB also
acknowledges the support of the Viola Vestal Coulter Foundation
through a research fellowship.

\end{document}